# NONCONTINOUS ADDITIVE ENTROPIES OF PARTITIONS

By Tomasz Sobieszek

*University of Łódź*[*]

We complete a description of **additive partition entropies**. This time we do not assume I to be continous. In the process we solve a 2-cocycle functional equation for certain subsets of convex cones.

**1. Introduction.** We assume all definitions and results from [3]. In the current follow-up paper we drop the assumption of the continuity of the additve partition entropy. Our aim is to prove a theorem analogical to Theorem 1 of [3]. As a by-product of our efforts we obtain that all symmetric solutions to a 2-cocycle functional equation for certain subsets of convex cones are coboundaries (Lemma 2, see also Remark 1).

**2. The results.** To begin with we need to extend Example 1 of [3].
By $\mathrm{End}(\mathbb{R})$ we shall mean the $\mathbb{Q}$-algebra of endomorphisms of the linear space $\mathbb{R}$ of reals over the field $\mathbb{Q}$ of rationals.

Example 1. *Consider a finitely-additive set function $\mathfrak{m}\colon \mathfrak{X} \to \mathrm{End}(\mathbb{R})$ vanishing on sets of $\mathsf{P}$-measure $0$. Then the mapping $\mathsf{L}_\mathfrak{m}\colon \mathfrak{A} \to \mathbb{R}$ defined by*

$$\mathsf{L}_\mathfrak{m}(\mathcal{A}) := \sum_{1\le i\le n} \mathfrak{m}(A_i)\left(\log \frac{1}{\mathsf{P}(A_i)}\right),$$

*where $\mathcal{A} = \langle A_1, \ldots, A_n\rangle$, is an additive partition entropy.*

Theorem 1. *Let $\mathrm{I}$ be an additive partition entropy. There exist an additive entropy $\mathsf{H}$, and a finitely-additive set function $\mathfrak{m}\colon \mathfrak{X} \to \mathrm{End}(\mathbb{R})$ vanishing on sets of $\mathsf{P}$-measure $0$ such that*

$$\mathrm{I} = \mathsf{H}_\mathsf{P} + \mathsf{L}_\mathfrak{m}.$$

Proof. According to Proposition 4 in [3], given any sets $V$, $W$ of the same measure $\mathsf{P}(V) = \mathsf{P}(W)$ one can define a mapping $\Delta(V, W) \in \mathrm{End}(\mathbb{R})$ as $\log\lambda \mapsto \Delta(V, W, \lambda)$. $\Delta$ satisfies the following conditions:

1. For any sets $U$, $V$ and $W$ of the same measure

$$\Delta(U, W) = \Delta(U, V) + \Delta(V, W).$$

---

[*] This paper is partially supported by Grant nr N N201 605840.

*AMS 2000 subject classifications:* Primary 94A17; secondary 60A10
*Keywords and phrases:* additive entropy, inset entropy, partition entropy, axioms of entropy, independent σ-algebras

2. For any sequences of disjoint sets $A, B, \dots, C$ and $A', B', \dots, C'$ which satisfy the equalities $P(A) = P(A'), \dots, P(C) = P(C')$ we have
$$\Delta(A + B + \dots + C, A' + B' + \dots + C') = \Delta(A, A') + \Delta(B, B') + \dots + \Delta(C, C').$$
3. Moreover by Remark 5 in [3], if $P(V \triangle V') = P(W \triangle W') = 0$, then
$$\Delta(V, W) = \Delta(V', W').$$

By analogy with Theorem 1 in [3] what we need is the following Lemma, which in some way can be regarded as a strengthening of Lemma 6 there

Lemma 1. *For any $\mathbb{Q}$-linear space $L$ and any $\Delta(V, W) \in L$, defined for sets $V$, $W$ of the same measure $P$ and satisfying conditions 1–3 above there is a finitely-additive set function $\mathfrak{m} \colon \mathcal{X} \to L$ such that $\mathfrak{m}(V) = 0$ whenever $P(V) = 0$ and also such that for any sets $V$ and $W$ satisfying $P(V) = P(W)$ we have*
$$\Delta(V, W) = \mathfrak{m}(W) - \mathfrak{m}(V).$$

Proof. To begin with, notice that there is a mapping $\mathfrak{m}_1 : \mathcal{X} \to L$ such that we have $\Delta(V, W) = \mathfrak{m}_1(W) - \mathfrak{m}_1(V)$ and $\mathfrak{m}_1(V) = 0$ when $P(V) = 0$. Indeed, for any $\theta \in [0, 1]$ choose an arbitrary $V_\theta$ such that $P(V_\theta) = \theta$ and set $\mathfrak{m}_1(V_\theta) = 0$. Next, for any $W$ of measure $P(W) = \theta$, set $\mathfrak{m}_1(W) = \Delta(V_\theta, W)$. Conditions 1 and 3 above give $\Delta(V, W) = \mathfrak{m}_1(W) - \mathfrak{m}_1(V)$ for any $V, W \in \mathcal{X}$, $P(V) = P(W)$.

By property 2 above the number
$$\mathfrak{m}_1(A + B + \dots + C) - \mathfrak{m}_1(A) - \mathfrak{m}_1(B) \cdots - \mathfrak{m}_1(C),$$
defined for disjoint $A, B, \dots, C \in \mathcal{X}$, depends solely on measures
$$P(A), P(B), \dots, P(C),$$
i.e. there is a function
$$f \colon \{(a, b, \dots, c) \in \mathbb{R}_+ \times \mathbb{R}_+ \times \dots \times \mathbb{R}_+ : a + b + \dots + c \le 1\} \to L$$
such that
$$\mathfrak{m}_1(A + B + \dots + C) - \mathfrak{m}_1(A) - \mathfrak{m}_1(B) \cdots - \mathfrak{m}_1(C) = f(P(A), P(B), \dots, P(C)). \tag{1}$$
We see that $f$ has the following properties
$$\begin{gathered} f(0, 0, \dots, 0) = 0, \\ f(a_{\sigma(1)}, a_{\sigma(2)}, \dots, a_{\sigma(n)}) = f(a_1, a_2, \dots, a_n), \end{gathered} \tag{2}$$
for any permutation $\sigma$, and also, as a direct consequence of (1),
$$\begin{gathered} f(a_1, \dots, a_k, b_1, \dots, b_l, \dots, c_1, \dots, c_m) = \\ f(a_1 + \dots + a_k, b_1 + \dots + b_l, \dots, c_1 + \dots + c_m) \\ + f(a_1, \dots, a_k) + f(b_1, \dots, b_l) + \dots + f(c_1, \dots, c_m). \end{gathered} \tag{3}$$

Suppose for now we can find a mapping $h\colon [0,1] \to L$ satisfying

$$f(a,b,\dots,c) = h(a+b+\dots+c) - h(a) - h(b) \,\cdots\, - h(c).$$

Then $h(0) = -f(0,0) = 0$. Moreover, setting $m(V) := m_1(V) - h(P(V))$ we obtain a new function $m : \mathcal{X} \mapsto L$. It follows straight from (1) that $m$ is finitely-additive. What is more, if $P(V) = P(W)$, then

$$m(W) - m(V) = m_1(W) - m_1(V) = \Delta(V,W).$$

The possibility of finding $h$ given (2) and (3) is a 'homological' fact which, being interesting in itself, we have separated as Lemma 2 below. □

Lemma 2. *Consider any $\mathbb{Q}$-linear spaces $K$ and $L$, a convex cone $P \subset K$ such that $P \cap -P = \{0\}$ and a set $M \subset P$ such that $P = \mathbb{Z}_+ M$ and let $a,b \in P$, $a+b \in M$ imply $a,b \in M$.*

*For any function*

$$f\colon \bigcup_{n\geq 1} \{(a_1,a_2,\dots,a_n) \in P\times P\times\dots\times P : a_1+a_2+\dots+a_n \in M\} \to L,$$

*the values of which do not depend on the permutation of arguments:*

$$f(a_{\sigma(1)},a_{\sigma(2)},\dots,a_{\sigma(n)}) = f(a_1,a_2,\dots,a_n), \tag{4}$$

*and which also satisfies*

$$\begin{aligned} f(a_1,\dots,a_k,b_1,\dots,b_l,\dots,c_1,\dots,c_m) &= \\ f(a_1+\dots+a_k,\, b_1+\dots+b_l,\dots,c_1+\dots+c_m)& \\ +f(a_1,\dots,a_k)+f(b_1,\dots,b_l)+\dots+f(c_1,\dots,c_m),& \end{aligned} \tag{5}$$

*there is a function $h\colon M \to L$ such that for $n \geq 1$*

$$f(a_1,a_2,\dots,a_n) = h(a_1+a_2+\dots+a_n) - h(a_1) - h(a_2) \,\cdots\, - h(a_n). \tag{6}$$

Proof. For a single argument, (5) gives $f(a) = 0$. From a repeated application of (5) we also get

$$f(a_1,\dots,a_n) = f(a_1+\dots+a_{n-1},a_n) + f(a_1+\dots+a_{n-2},a_{n-1}) + \dots + f(a_1,a_2).$$

Thus (6) will follow once we show that $h$ satisfies

$$f(a,b) = h(a+b) - h(a) - h(b). \tag{7}$$

Observe that for any $s \in M$, $\alpha \in \mathbb{Q}$, $\alpha \leq 1$ we have $\alpha s, (1-\alpha)s \in P$, $\alpha s + (1-\alpha)s = s \in M$, thus $\alpha s \in M$.

In order to shorten notation we shall write $a : k$ for $\overbrace{a,\dots,a}^{k}$. From (5) and in particular from $f(0) = 0$ we find that there is a constant $z \in L$ such that

$f(0:n) = (n-1)z$, $n \geq 1$. By (7) any solution $h$ must have $f(a, 0) + h(0) = 0$ and $h(0) = -z$. [1]

For any $s \in M\setminus\{0\}$ and for the ray $S = \mathbb{Q}_+ s$, there exists a solution $h_S \colon S\cap M \to L$ to equation (7) restricted to $a, b, a+b \in S \cap M$, with arbitrary value $h_S(s) \in L$. Indeed, begin by writing $h_S(0) = -z$, choose arbitrarily $h_S(s) \in L$ and then extend

$$h_S(\tfrac{p}{q}s) := \tfrac{p}{q}\left[h_S(s) - f(\tfrac{s}{q} : q)\right] + f(\tfrac{s}{q} : p), \tag{8}$$

for any positive integers $p$, $q$ such that $\frac{p}{q}s \in M$. The function $h_S$ is well-defined since for any $n \geq 2$, equality (5) gives us

$$\begin{aligned} f(\tfrac{s}{qn} : pn) - \tfrac{p}{q}f(\tfrac{s}{qn} : qn) &= \\ &= \left[f(\tfrac{s}{q} : p) + pf(\tfrac{s}{qn} : n)\right] - \tfrac{p}{q}\left[f(\tfrac{s}{q} : q) + qf(\tfrac{s}{qn} : n)\right] \\ &= f(\tfrac{s}{q} : p) - \tfrac{p}{q}f(\tfrac{s}{q} : q). \end{aligned}$$

Moreover, $h_S$ satisfies (7). In fact, given positive $k, l$, from

$$f(\tfrac{s}{q} : k+l) = f(\tfrac{k}{q}s, \tfrac{l}{q}s) + f(\tfrac{s}{q} : k) + f(\tfrac{s}{q} : l),$$

we have by (8)

$$h_S(\tfrac{k+l}{q}s) = f(\tfrac{k}{q}s, \tfrac{l}{q}s) + h_S(\tfrac{k}{q}s) + h_S(\tfrac{l}{q}s),$$

Which is also true if any of $k, l$ is $0$. Since equality (7) implies formula (8) there is only one such solution $h_S$ having a fixed value at a given nonzero point $s \in S \cap M$.

Suppose that for some convex cone $C \subset P$ we are given a solution $h_1 \colon C\cap M \to L$ to equation (7) restricted to $a, b, a+b \in C \cap M$. As we shall presently show, for any $s \in M \setminus C$ there is an extension $h\colon (C \vee \mathbb{Q}_+ s) \cap M \to L$ which continues to satisfy equation (7) with $a, b, a+b \in (C \vee \mathbb{Q}_+ s) \cap M$.

Suppose first that $s \notin \mathrm{Span}(C)$. Any element of $(C \vee \mathbb{Q}_+ s) \cap M$ can be then uniquely represented as $a + b \in M$, with $a \in C$, $b \in \mathbb{Q}_+ s$. Define for such an element

$$h(a+b) := h_1(a) + h_S(b) + f(a, b), \tag{9}$$

where $h_S$ is any solution to (7) on $\mathbb{Q}_+ s \cap M$. In particular, $h(a) = h_1(a)$, $h(b) = h_S(b)$. What is more, for any $a, a' \in C$ and $b, b' \in \mathbb{Q}_+ s$, such that $a + a' + b + b' \in M$ we obtain

$$\begin{aligned} h(a+b+a'+b') &= \\ &= h(a+a') + h(b+b') + f(a+a', b+b') && \text{(by (9))} \\ &= h(a) + h(a') + h(b) + h(b') + f(a, a', b, b') && \text{(by (7), (5))} \\ &= h(a+b) + h(a'+b') + f(a+b, a'+b'). && \text{(by (9), (5))} \end{aligned}$$

---

[1] By replacing $f(a_1, a_2, \ldots, a_n)$ with $f(a_1, a_2, \ldots, a_n) - (n-1)z$ and $h(a)$ with $h(a) + z$ we could assume that $f(0, 0, \ldots, 0) = 0$. Then we would have $f(a, 0) = 0$ and $h(0) = 0$.

Suppose now that $s \in \mathrm{Span}(C)$. In this case there are $r \in C$, $\alpha_0 \in \mathbb{Q}_+$ with $r + s \in C$ and $\alpha_0(r+s) \in M$. Consider a solution $h_S$ to equation (7) on $\mathbb{Q}_+ s \cap M$ with $h_S(\alpha_0 s)$ chosen in such a way so as to satisfy

$$h_1(\alpha_0(r+s)) = h_1(\alpha_0 r) + h_S(\alpha_0 s) + f(\alpha_0 r, \alpha_0 s). \tag{10}$$

Now, observe that grouping $\alpha r, \beta r, \alpha s, \beta s$ in two different ways gives by (5)

$$\begin{aligned} f(\alpha r, \beta r, \alpha s, \beta s) &= \\ &= f(\alpha(r+s), \beta(r+s)) + f(\alpha r, \alpha s) + f(\beta r, \beta s) \\ &= f((\alpha+\beta) r, (\alpha+\beta) s) + f(\alpha s, \beta s) + f(\alpha r, \beta r) \end{aligned} \tag{11}$$

whenever $(\alpha+\beta)(r+s) \in M$. It follows that the mapping $H_S$ defined for such $\alpha$s that $\alpha(r+s) \in M$ by

$$H_S(\alpha s) := h_1(\alpha(r+s)) - h_1(\alpha r) - f(\alpha r, \alpha s)$$

satisfies

$$\begin{aligned} H_S(\alpha s + \beta s) - H_S(\alpha s) - H_S(\beta s) &= \\ &= f(\alpha(r+s), \beta(r+s)) - f(\alpha r, \beta r) && \big(\text{by (7) for } h_1\big) \\ &\quad - f((\alpha+\beta) r, (\alpha+\beta) s) + f(\alpha r, \alpha s) + f(\beta r, \beta s) \\ &= f(\alpha s, \beta s) && \big(\text{by (11)}\big) \end{aligned}$$

i.e. condition (7) on the set $M' = \{\alpha s : \alpha(r+s) \in M, \alpha \in \mathbb{Q}_+\}$. Since $M'$ plays the role of $M$ for the ray $\mathbb{Q}_+ s$ and $H_S(\alpha_0) = h_S(\alpha_0)$, $H_S$ is a restriction of $h_S$ to $M'$. This shows that

$$h_1(\alpha(r+s)) = h_1(\alpha r) + h_S(\alpha s) + f(\alpha r, \alpha s), \tag{12}$$

for any $\alpha \in \mathbb{Q}_+$, $\alpha(r+s) \in M$. Define $h$ by

$$h(a+b) := h_1(a) + h_S(b) + f(a, b),$$

for $a \in C$, $b \in \mathbb{Q}_+ s$, $a + b \in M$. It remains to check that $h$ is well-defined, and then to proceed like in the previous case with $s \in \mathrm{Span}(C)$. To this end, let $a' - a = b - b' = \alpha s$, where $a' \in C$, $b' \in \mathbb{Q}_+ s$. We can assume that $\alpha \geq 0$. Then for $\alpha r := a$ equality (12) takes the following form:

$$h_1(a') = h_1(a) + h_S(b - b') + f(a, b - b').$$

Additionaly $f(a, b-b') + f(a', b') = f(a, b-b', b') = f(b-b', b') + f(a, b)$. Therefore

$$\begin{aligned} h_1(a') + h_S(b') + f(a', b') &= \\ &= h_1(a) + h_S(b') + h_S(b-b') + f(a, b-b') + f(a', b') \\ &= h_1(a) + h_S(b') + h_S(b-b') + f(b-b', b') + f(a, b) \\ &= h_1(a) + h_S(b) + f(a, b). && \big(\text{by (7) for } h_S\big) \end{aligned}$$

The existence of $\mathfrak{h}$ satisfying (7) for $a, b, a+b \in (C \vee \mathbb{Q}_+ s) \cap M$ is thus proved.

Consider the family $\mathcal{H}$ of the solutions $\mathfrak{h}$ to equation (7), defined on the intersections of shape $C \cap M$, where $C \subset P$ is a convex cone. $\mathcal{H}$ is trivially nonempty and inductive, (with respect to the partial order defined by $\mathfrak{h}_1 \preceq \mathfrak{h}_2 \iff \mathfrak{h}_1 \subset \mathfrak{h}_2$). By Kuratowski-Zorn Lemma the family $\mathcal{H}$ has a maximal element $\mathfrak{h}_C \colon C \cap M \to L$. We shall show by contraposition that $M \subset C$. Indeed, assume that we can find $s \in M$, $s \notin C$. It allows us, according to previous considerations, to define a mapping $\mathfrak{h} \in \mathcal{H}$ on the set $(C \vee \mathbb{Q}_+ s) \cap M$ which extends $\mathfrak{h}_C$. □

This concludes the proof of Theorem 1. □

It should be possible to derive a version of Lemma 2 with considerably weaker assumptions on the shape of $M$. It seems, however, that it would make the proof longer still.

The remark below makes it easy to spot a relation between the function $f$ of the last Lemma and other similar notions that appear elsewere. In fact, it shows that the Lemma is a variant of a result by Erdős [2], that a symmetric 2-cocylcle is a coboundary, see [1].

Remark 1. *In the assumptions of Lemma 2 it suffices to define $f$ for two arguments and replace (4), (5) with the conditions*

$$\begin{aligned} f(a,b) &= f(b,a), \\ f(a,b+c) + f(b,c) &= f(a+b,c) + f(a,b). \end{aligned} \tag{13}$$

In fact, we extend the definition of $f$ by setting:

$$\begin{aligned} f(a) &:= 0 \\ f(a,b,\dots,c) &:= f(a+b,\dots,c) + f(a,b), \end{aligned}$$

for any $a, b, \dots, c \in P$ such that $a + b + \cdots + c \in M$.

We have to show (4) and (5). In fact, the former is obvious when there are at most three arguments. For $n > 3$ arguments we use induction. By the definition of $f$ it follows, that we can permute the last $n-2$ arguments without changing the value of $f$. Since, by definition of $f$ for the sequences $(a, b, c, \dots, d)$, $(a+b, c, \dots, d)$ and $(a, b, c)$ we have

$$f(a,b,c,\dots,d) = f(a+b+c,\dots,d) + f(a,b,c),$$

we can also permute the first 3 arguments. This shows (4). Coming over to (5) it suffices to show it in case when one of the numbers $k, l, \dots, m$ is greater than 1. We can assume that $k \geq 2$. Then the sought-after equality follows from the same equality for the shorter sequence

$$(a_1 + a_2, \dots, a_k, b_1, \dots, b_l, \dots, c_1, \dots, c_m)$$

and the definitions of $f$ for both sequences $(a_1, \dots, a_k)$ and

$$(a_1, \dots, a_k, b_1, \dots, b_l, \dots, c_1, \dots, c_m).$$

A reasoning similar to that of Remark 1 in [3] gets us the following

Remark 2. *The additive partition entropy of form* $L_{\mathfrak{m}}$*, for a finitely-additive set function* $\mathfrak{m}\colon \mathcal{X} \to \mathrm{End}(\mathbb{R})$ *vanishing on sets of* $P$*-measure* $0$*, depends solely on measures of atoms exactly when there is an* $\alpha \in \mathrm{End}(\mathbb{R})$ *such that* $\mathfrak{m}(A) = \alpha(P(A))$ *for any* $A \in \mathcal{X}$*.*

**References.**

[1] Ebanks, B. R. (2007). General solution of the 2-cocycle functional equation on solvable groups *Aequationes Math.* **73** 260–279.

[2] Erdős, P. (1959). A remark on the paper 'On some functional equations' by S. Kurepa *Glas. Mat.-Fiz. Astron. Ser. II* **14** 3–5.

[3] Paszkiewicz, A., and Sobieszek, T. *Additive partition entropies* (preprint)

Faculty of Mathematics and Computer Science
University of Łódź
ul. Banacha 22, 90-238 Łódź
Poland
E-mail: sobieszek@math.uni.lodz.pl
URL: http://sobieszek.co.cc